\documentclass[preprint,12pt,authoryear]{elsarticle}

\usepackage{natbib}
\usepackage{amsmath,amssymb,amsfonts}
\usepackage{algorithmic}
\usepackage{graphicx}
\graphicspath{{figures/}}
\usepackage{textcomp}
\usepackage{xcolor}

\usepackage{blindtext}
\usepackage{balance}
\usepackage{multirow}
\usepackage{hyperref}
\usepackage{longtable}
\usepackage{enumitem}
\usepackage[normalem]{ulem}
\useunder{\uline}{\ul}{}
\usepackage{lscape}
\usepackage{float}
\usepackage{booktabs}
\usepackage{xcolor}
\usepackage{mdframed}

\definecolor{lightgray}{gray}{0.95}

\newmdenv[
  backgroundcolor=lightgray,
  linecolor=gray,
  linewidth=1pt,
  roundcorner=5pt,
  skipabove=10pt,
  skipbelow=10pt
]{summarybox}

\journal{Information and Software Technology}

\begin{document}

\begin{frontmatter}

\title{Recommendations for Efficient and Responsible LLM Adoption within Industrial Software Development}

\author[label1]{Krishna Ronanki}
\author[label1]{Beatriz Cabrero-Daniel}
\author[label2]{Tomas Herda}
\author[label2]{Stefan Sitkovich}
\author[label1]{Jennifer Horkoff}
\author[label1]{Christian Berger}
\affiliation[label1]{organization={Chalmers University of Technology | University of Gothenburg},
 city={Gothenburg},
country={Sweden}}
\affiliation[label2]{organization={Austrian Post},
 city={Vienna},
 country={Austria}}

\begin{abstract}
\textbf{Context:} Large language models (LLMs) are observed to have a significant positive impact on various software engineering (SE) activities. With improved accessibility, the adoption of powerful LLMs in industry has surged recently. However, there is a lack of actionable best practices for the efficient and responsible adoption of LLMs within industrial software settings.
\textbf{Objectives:} We developed seven actionable recommendations to address this research gap.
\textbf{Methods:} We conducted a multi-case study with three organisations that use LLMs within their SE activities and synthesised seven recommendations through qualitative thematic analysis. We conducted a complementary online survey with software practitioners from various industries to evaluate the perceived relevance of our recommendations.
\textbf{Results:} Our results and recommendations focus on (i) users' preference to use LLMs as AI assistants, (ii) the importance of relevant stakeholders' satisfaction in the LLM-output evaluation, (iii) scoping the applicability of LLMs within SE tasks, (iv) the effect of LLMs on SE workflows, (v) the necessity and directions for developing human oversight mechanisms, and (vi) the necessary skills for practitioners for leveraging LLMs within SE. The online survey indicates a high level of agreement from the participants regarding the perceived relevance of the recommendations.
\textbf{Conclusion:} We outline future research directions, including mapping the seven recommendations to the principles of the EU AI Act (AIA) in order to examine how they relate to the current regulatory compliance frameworks.

\end{abstract}


\end{frontmatter}

%%\linenumbers

%% main text
\section{Introduction} \label{sec:Intro}

Many state-of-the-art LLMs are demonstrating increasingly impressive capabilities in performing a wide range of tasks~\citep{brown2020language}, including tasks within software engineering (SE), and are observed to increase user productivity~\citep{peng2023impact}. The adoption of LLMs in industrial SE settings is observed to be growing thanks to easy-to-use interfaces that make the LLMs more accessible to practitioners of all backgrounds~\citep{white2023chatgpt}.

Despite their ease of use, there are a few impediments in using LLMs for SE in practice such as confidential data privacy, extrinsic hallucinations~\citep{bang2023multitask}, and the lack of best practices~\citep{10449667}. Third party LLM-based solutions such as Azure OpenAI services~\citep{azure}, Microsoft Copilot~\citep{copilot}, GitHub Copilot~\citep{ghcp}, ChatGPT enterprise~\citep{gptent} aim to mitigate some of these issues.

However, that still leaves the shortage of best practices for using LLMs in industrial SE as an open challenge. This points toward the need for more empirical research on LLM-based SE~\citep{10449667}. To address this gap, we define the research problem addressed in this paper as follows:

\begin{summarybox}
\textbf{Users and organisations lack actionable recommendations to leverage LLMs to assist them in their SE activities in an AIA-compliant manner.}
\end{summarybox}

Although studies such as \cite{10449667} and \cite{10.1145/3695988} reinforce the benefits and highlight the challenges of leveraging LLMs within SE, the insights presented are based on secondary evidence-based studies~\citep{kitchenham2004evidence} or controlled empirical experiments~\citep{wohlin2012experimentation} that do not fully capture the nuances of adopting LLM in industrial settings. Hence, we see a clear need for an empirical study that focuses on directly capturing practitioner's experiences regarding the adoption of LLMs within industrial SE environments. With the software industry's increasing interest in leveraging LLMs in their processes, more empirical evidence collected from real-world environments is required to devise and understand the mechanisms that facilitate efficient and responsible adoption of LLMs in SE~\citep{10449667}.

To that extent, we conducted an interview-based multi-case study with three organisations to gather empirical evidence regarding the adoption of LLM-based tools and services to assist users in their SE activities. We conducted a total of fifteen interviews with eighteen different participants across the three organisations. We collected qualitative data regarding their insights on their use cases, motivation, intentions and process of integrating LLMs in various SE activities, as well as the advantages and limitations of adopting LLMs within industrial settings. Based on these findings, we developed seven recommendations for the adoption of LLMs for SE within industrial settings.

However, findings of a multi-case study are limited to the context of the case companies under investigation and the recommendations are potentially not directly generalisable to broader contexts of adopting LLMs for SE activities. Therefore, we complemented our multi-case study results with a broader survey with software practitioners to assess their level of agreement with the recommendations that we synthesised from our study. We performed this survey to investigate the perceived relevance of our recommendations, as the software practitioners we surveyed were sampled from various industries, who, self-reportedly, have experienced LLM adoption within various SE tasks under different contexts. We conclude our study with a post hoc mapping of the validated recommendations with the seven key trustworthy AI principles proposed by the European Commission's High Level Expert Group on Artificial Intelligence (AI HLEG). Based on the results of this process, the contributions of this study are as follows:

\begin{itemize}[noitemsep]
 \item Findings surrounding the benefits, concerns, challenges, limitations, and experiential learnings of LLM-assisted SE in practice.
 \item Seven recommendations for the efficient and responsible adoption of LLMs in industrial SE activities.
 \item Assessing the perceived relevance of the recommendations beyond the context of the case companies.
\end{itemize}
 
The rest of the paper is organised as follows. Section~\ref{sec:bg} covers the background concepts relevant to our study. Section~\ref{sec:rw} reviews related works that inform our research motivation and provide a basis for comparing the results of our study. Section~\ref{sec:methods} describes our research design. Section~\ref{sec:mcs} presents the results and analysis of the multi-case study we conducted. Section~\ref{sec:recs} presents the seven proposed recommendations. Section~\ref{sec:survey} presents the practitioner's perception of the proposed recommendations. We present our post hoc mapping of the validated recommendations with the seven AI HLEG's trustworthy AI principles in Section~\ref{sec:tai}. We discuss the implications and the validity threats to our results in Section~\ref{sec:discussion}. Finally, we conclude our study in Section~\ref{sec:conclusion} and present potential areas for further research.

\section{Background} \label{sec:bg}

LLMs, often built upon deep learning techniques like transformers, can produce useful natural language outputs. This led to them being employed in various language-related tasks such as text generation~\citep{goyal2022news}, question answering~\citep{nakano2021webgpt}, translation~\citep{brown2020language}, summarisation~\citep{xie2023survey}, and sentiment analysis~\citep{kheiri2023sentimentgpt}. The application of LLMs for automating or assisting users within certain SE practices is an emerging area of interest. This is indicated by the studies conducted by \cite{10449667} and \cite{10.1145/3695988} who, respectively, presented a survey and systematic literature review on the application of LLMs within SE.

\subsection{Retrieval Augmented Generation}

While LLMs have achieved remarkable success, they still encounter some substantial limitations. These are particularly evident in tasks that are specialised or require extensive knowledge~\citep{kandpal2023large}. Notably, they tend to generate misleading or false information termed as ``hallucinations''~\citep{zhang2023siren} when dealing with inquiries that exceed their training data or necessitate up-to-date information.

To address this limitation, recent research has explored the integration of both parametric and non-parametric memory into LLMs. Parametric memory refers to the learned parameters within the model, while non-parametric memory typically involves accessing external knowledge sources such as large text corpora or databases. One approach to combining these types of memory is through a method called Retrieval Augmented Generation (RAG). RAG endows pre-trained parametric-memory generation models, such as transformers, with non-parametric memory by incorporating a dense vector index of external knowledge. This integration is achieved through a general-purpose fine-tuning approach, allowing the model to access extensive knowledge during inference without additional training~\citep{lewis2021retrievalaugmented}.

\subsection{Trustworthy AI Guidelines}

In order to achieve trustworthy AI in practice, the AI HLEG proposed seven key principles that must be continuously assessed and managed throughout the entire lifecycle of an AI system: \textit{Human Agency and Oversight}, \textit{Technical Robustness and Safety}, \textit{Privacy and Data Governance}, \textit{Transparency}, \textit{Diversity, Non-discrimination, and Fairness}, \textit{Societal and Environmental Well-being}, and \textit{Accountability}~\citep{aihleg,AIA}. These principles are not legally binding under the AIA but are referenced in Recital (27) as voluntary guidance for developing best practices and standards that foster trustworthy, human-centric AI~\citep{AIA}.

It is beneficial to ensure all actors involved in LLM-assisted SE tasks or processes comply with the trustworthy AI principles. According to the AIA, ``any natural or legal person, including a public authority, agency or other body, using an AI system under its authority, except where the AI system is used in the course of a personal non-professional activity'', is referred to as a deployer~\citep{AIA}. Under this definition, employees of software organisations who participate in the design, development, deployment and maintenance of a software system can be categorised as deployers if they use the assistance of an LLM. As such, we hypothesise that adhering to AI HLEG's seven principles during the adoption and integration of LLMs within SE workflows can aid the deployers in aiming for trustworthiness.

\section{Related Work} \label{sec:rw} 

In this section, we provide a brief summary of the state-of-the-art literature relevant to our study, particularly looking at empirical studies applying LLMs in SE contexts. 

\cite{Pereira2024} conducted an industrial case study with a large media group that has recently begun to adopt OpenAI ChatGPT and GitHub Copilot for software development activities and provide early insights into potential benefits and concerns in the form of eight initial lessons learnt. These eight lessons are centred around aspects such as generative AI's impact on a user's learning, improving unit testing, the importance of context and the user's domain expertise when using LLMs and other potential benefits and challenges surrounding the usage of LLMs within SE.

Similarly, \cite{10.1007/978-3-031-42622-3_49} conducted a case study, focusing on how employees in SE perceive the collaboration with AI-powered chatbots such as ChatGPT. They identified fourteen distinct insights into the perceived collaboration with AI-powered chatbots in the case company's software development context: Developing a general understanding on how AI works, possessing expertise on the task or topic to which the AI agent is assigned, interpreting and evaluating AI agent's outputs context-specifically, determining the division of tasks between the AI and oneself, dealing with the AI agent's outputs in a reflective manner, complying with data protection rules, considering the AI agent as an enabler, considering the AI agent as a sort of a virtual colleague, being able to adapt and be open for change and innovation, feeling confident to work with new and unfamiliar technologies, complying with ethical and moral standards, appreciating the AI agent's support, engaging oneself in a constant discourse with the AI agent, and expressing oneself comprehensibly towards the AI agent.

\cite{WANG2025104113} conducted a mixed methods field study to incorporate LLMs into the vulnerability remediation process effectively. As part of their study, they design, implement, and empirically validate an LLM-supported collaborative vulnerability remediation process. The lessons learnt from this study are centred around incorporating LLMs into practical processes, facilitating collaboration among all associated stakeholders, reshaping LLMs' roles according to task complexity, and how to approach the short-term side effects of improved user engagement facilitated by LLMs.

While these studies offer valuable early perspectives drawn from individual organisational contexts, they primarily adopt a descriptive stance, capturing lessons learnt and practitioner perceptions regarding the adoption and integration of LLMs in SE. However, as organisations increasingly explore operational adoption, there remains a need for prescriptive, actionable guidance that extend beyond descriptive observations to support efficient and responsible adoption in practice. Addressing this gap, our study developed and validated a set of recommendations to guide the adoption of LLMs within SE organisations. We also compare and contrast the contributions of our study with the lessons presented by \cite{Pereira2024}, \cite{10.1007/978-3-031-42622-3_49}, and \cite{WANG2025104113} in Section~\ref{sec:discussion}.

\section{Methodology} \label{sec:methods}

\subsection{Overview} \label{subsec:overview}

\begin{figure*}[ht!]
 \centering
\includegraphics[width=\linewidth]{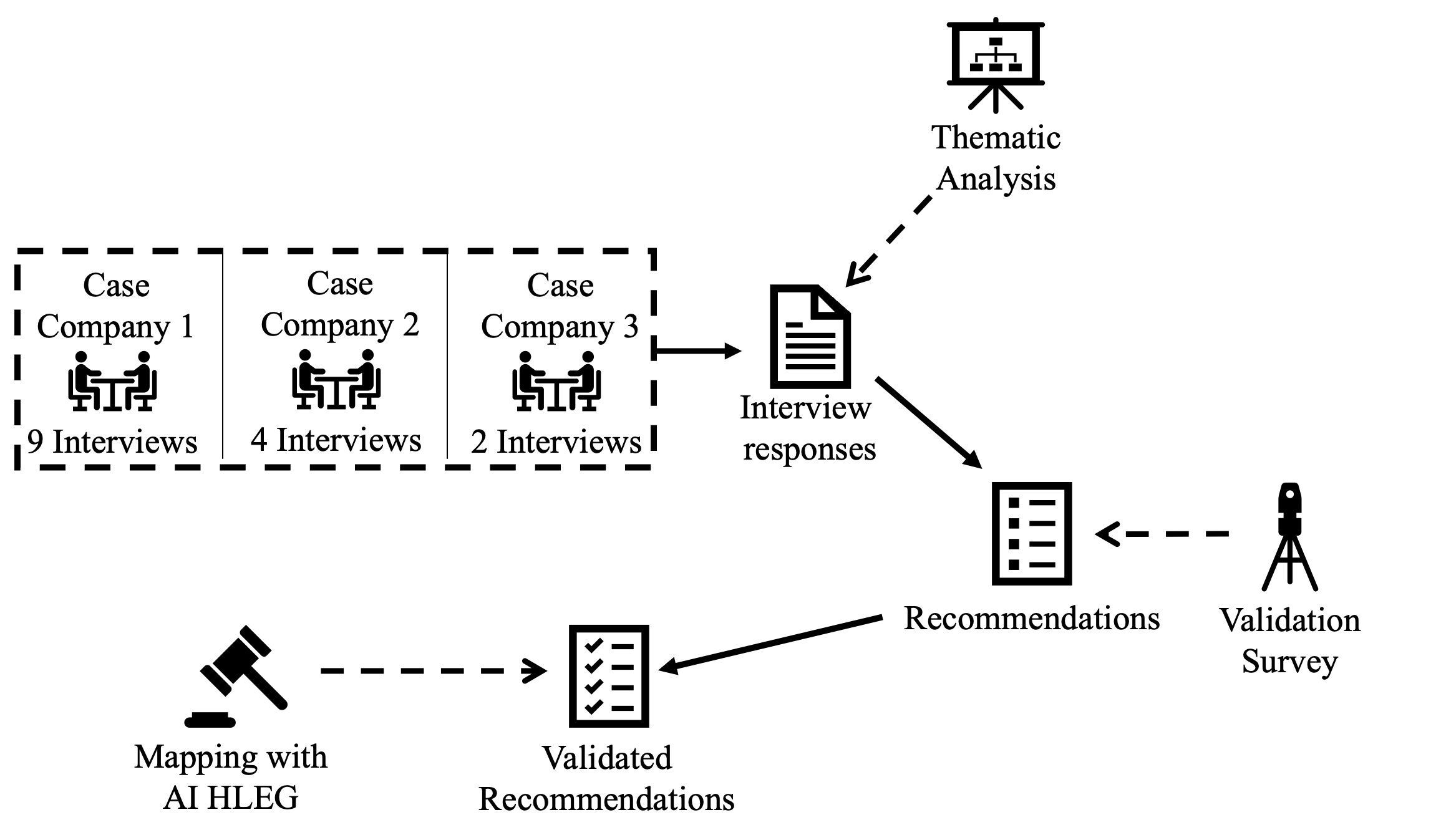}
\caption{Methodology. The dotted arrows represent the means of data collection or analysis while the solid arrows point to the outcome of the applied method.}
\label{fig:fig-1}
\end{figure*}

Figure~\ref{fig:fig-1} is a visual representation of this study's design. We employed a mixed methods approach as part of our research design. We provide an overview of our study design in this section, with additional details regarding the description of the case companies along with the data collection and analysis activities in subsections~\ref{subsec:casedes}, \ref{subsec:dcol}, \ref{subsec:dan}, \ref{subsec:survey}, and \ref{subsec:mapping} respectively.

We formulated the following research questions (\textbf{RQs}) to obtain our study's contributions:

\begin{summarybox}
\textbf{RQ1.~What are the benefits, challenges and experiential learnings of practitioners concerning LLMs' assistance in their SE activities?}
\vspace{5pt}
\end{summarybox}

\noindent \textbf{Justification:} The insights from prior studies have mainly been derived from individual case studies conducted in specific organisational contexts. This leaves a need for further empirical work that examines whether similar benefits, challenges, and experiential learnings can be observed across multiple cases. Building on these findings, our RQ1 examines practitioners' reported benefits, challenges, and experiential learnings from using LLMs in SE activities.

\begin{summarybox}

\textbf{RQ2.~What are recommendations for the adoption of LLMs' assistance for SE activities?}
\vspace{5pt}
\end{summarybox}

\noindent \textbf{Justification:} Existing work mainly reports lessons learnt and practitioner perceptions \citep{Pereira2024,10.1007/978-3-031-42622-3_49,WANG2025104113}. While these studies improve understanding of LLM use in practice, they provide limited explicit guidance for adoption. Based on the experiences captured in RQ1, our RQ2 focuses on deriving recommendations for adopting LLM assistance in SE activities.

\begin{summarybox}
\textbf{RQ3.~What is the practitioners' view on the perceived relevance of the proposed recommendations?}
\vspace{5pt}
\end{summarybox}

\noindent \textbf{Justification:} Prior studies suggest that the use of LLMs in SE is shaped by context, practitioner expertise, and task characteristics \citep{10.1007/978-3-031-42622-3_49,WANG2025104113}. For this reason, proposed recommendations should also be examined from the perspective of practitioners. Our RQ3 therefore investigates practitioners' views on the perceived relevance of the proposed recommendations.

We conducted the interview-based multi-case study following the guidelines for conducting and reporting case study research in SE by \cite{runeson2009guidelines} to answer \textbf{RQ1}. We conducted a total of fifteen interviews with eighteen participants across three cases. Case 1 had five individual interviews and four group interviews. The group interviews were conducted with three people in each. Case 2 had four individual interviews while case 3 had two individual interviews.

We performed reflexive thematic analysis to answer \textbf{RQ2}. We followed the guidelines for conducting thematic synthesis proposed by \cite{Clarke2014} on the data collected from all fifteen interviews, which resulted in six primary themes having 13 codes. Based on the results of the thematic analysis, we synthesised seven recommendations. This was followed by a survey to assess the perceived relevance of these recommendations beyond the contexts of the case companies to answer \textbf{RQ3}. Together, the multi-case study and the survey addressed the overarching goal of synthesising and validating recommendations for the efficient and responsible adoption of LLMs within SE.

After answering the \textbf{RQs}, we conducted a post-hoc mapping of our seven recommendations to the seven key principles for trustworthy AI outlined by the AI HLEG. This mapping was done to assess how the recommendations relate to the EU AI Act (AIA).

The multi-case study provided in-depth findings by focusing on unique aspects of real-world adoption of LLMs. The online survey broadened the scope by involving practitioners from diverse organisational and contextual backgrounds. This approach validated the applicability and relevance of the recommendations across a variety of scenarios. The multi-case study acted as an exploratory phase, uncovering key themes and actionable recommendations grounded in the experiences of the case companies. The survey was conducted as a validation phase, assessing the perceived relevance of the recommendations across a diverse group of practitioners, reinforcing their reliability.

\subsection{Case Descriptions} \label{subsec:casedes}

\subsubsection{Case 1}

Case company 1 is an Europe-headquartered international postal and logistics service provider, whose main business activities include shipping and delivering letter mail, advertising mail, print publications and parcels. The case company's IT division is responsible for the design, development, deployment, maintenance and operation of a suite of software products and services that support the case company's logistical operations. The roles of the participants, the LLM-based tools used and the LLM-assisted SE use cases within case company 1 are presented in Table~\ref{tab:Table-1}.

\begin{table}[ht!]
\begin{tabular}{|p{4.5cm}|p{4cm}|p{4cm}|}
\hline
\multicolumn{1}{|c|}{\textbf{Roles}} & \multicolumn{1}{c|}{\textbf{Use cases}}  & \multicolumn{1}{c|}{\textbf{LLM tools used}} \\ \hline
Requirements  Engineer, Scrum Master, Product Owner, SE Team Lead & User Stories generation, Stakeholder Analysis Artefact generation & GitHub Copilot, Microsoft Copilot, RAG POC tool \\ \hline
\end{tabular}
\caption{Participant roles, LLM-based tools and LLM-assisted use cases of case 1}
\label{tab:Table-1}
\end{table}

Case company 1 had their own proof-of-concept RAG implementation called \textbf{RAG POC tool}, which was also tested within their workshop for generating the stakeholder analysis artefact. The stakeholder analysis artefact (case company specific artefact) serves to identify stakeholders (a person or an organisation who influences a system's requirements or who is impacted by that system) and assigns them clear responsibilities. The \textbf{RAG POC tool} was developed within case company 1 using Microsoft Azure Open AI services and a standard RAG architecture~\citep{Kumar2025}.

\subsubsection{Case 2}

Case company 2 is a Europe-headquartered multinational organisation in telecommunications and networking. It develops and deploys mobile network technologies, and provides infrastructure, software, and services for global connectivity. Case company 2 has an internal LLM tool, which shall remain unnamed within this article for confidentiality reasons. The internal LLM tool was developed in-house and is based on a RAG implementation. This RAG implementation has access to over 2000 company specific documents, and is primarily used for the information search and retrieval use case.

We conducted four individual semi-structured interviews with the participants from this case company. The tools used and the LLM-based use cases at this case company are presented in Table~\ref{tab:Table-2}.

\begin{table}[ht!]
\begin{tabular}{|p{4.5cm}|p{5cm}|p{3cm}|}
\hline
\multicolumn{1}{|c|}{\textbf{Roles}} & \multicolumn{1}{c|}{\textbf{Use cases}}  & \multicolumn{1}{c|}{\textbf{LLM tools used}} \\ \hline
System Manager, Product Owner, Developer and Scrum Master, Customer Product Information (CPI) Architect & Software documentation, coding assistance, information search and retrieval, brainstorming and ideation, scrum activity support & GitHub Copilot, Codeium, Internal LLM tool, ChatGPT \\ \hline
\end{tabular}
\caption{Participant roles, LLM-based tools and LLM-assisted use cases of case 2}
\label{tab:Table-2}
\end{table}

\subsubsection{Case 3}

Case company 3 is a Europe-headquartered multinational organisation in transportation and infrastructure. It develops and delivers digital services, focusing on connectivity, data analytics, and fleet management to enhance vehicle performance and operational efficiency.

We conducted two individual semi-structured interviews with the participants from this case company. The tools used and the LLM-based use cases at this case company are presented in Table~\ref{tab:Table-3}.

\begin{table}[ht!]
\begin{tabular}{|p{4cm}|p{5.5cm}|p{3cm}|}
\hline
\multicolumn{1}{|c|}{\textbf{Roles}} & \multicolumn{1}{c|}{\textbf{Use cases}}  & \multicolumn{1}{c|}{\textbf{LLM tools used}} \\ \hline
Developer and Scrum Master, Solution Architect & Coding Assistance, Unit Tests Generation, Cloud Infrastructure Management, Documentation  & GitHub Copilot, Amazon Q developer, ChatGPT \\ \hline
\end{tabular}
\caption{Participant roles, LLM-based tools and LLM-assisted use cases of case 3}
\label{tab:Table-3}
\end{table}

\subsection{Data Collection} \label{subsec:dcol}

We employed semi-structured interviews~\citep{1130282271686871424} as the method of data collection for the multi-case study. All interviews were conducted while following an ethical interview checklist~\citep{8870192}. They were audio-recorded with participant consent and transcribed verbatim using the the video conferencing tool's built-in record and transcribe feature. Due to scarcity of industry practitioners' time, no pilot interviews were conducted at any of the case companies.

The interview participants were selected using a purposive sampling approach~\citep{10.1007/s10664-021-10072-8}, as this method is well-suited for exploratory qualitative studies where participants are chosen to provide relevant insights on the phenomenon being investigated. Our primary inclusion criterion was that participants actively use LLMs in at least one of their regular SE use cases. Our intention behind this purposive approach emphasised that the selected participants had direct, practical experience with integrating LLMs into their SE workflows, aligning with the study's objective of exploring real-world applications and perceptions of LLM adoption within SE tasks. The exception to this criterion was case 1, where they were still in the pilot phase of adopting LLMs and started the adoption process in parallel with the case study and the participants did not have much experience using LLMs in their use cases.

We used different interview questionnaires for case 1 and case 2 \& 3. This is because of the difference in their LLM adoption contexts. The interview participants from case 2 and case 3 have been using LLMs within some of their software development activities for a year while case company 1 had just began to pilot test LLM-based use cases as of the date of data collection. We used different questionnaires to account for this difference when collecting the data. The two sets of interview questionnaires for the three cases are provided within an online repository\footnote{Zenodo repository containing the supplementary material, i.e., interview and survey questionnaires: \url{https://doi.org/10.5281/zenodo.15754264}}.

We collected data from case company 1 with two rounds of interviews. The first round of interviews (round-1 interviews) was individual and conducted with five people. The aim of these interviews was to understand participants' motivations for adopting LLMs, their evaluation strategies, expected improvements, and key concerns, including ethical and legal considerations for LLM-assisted RE workflows. The purpose was to gain insights into the feasibility and impact of LLM-based techniques, informing best practices for their adoption and optimisation in RE workflows.

A pilot testing of LLM-assisted RE workflows was conducted during an internal ``workshop'' hosted and organised by the case company. This internal workshop allowed users to experiment with LLM-assisted RE workflows and gain practical insights on how LLM could be implemented within their RE use cases. However, no data was collected during workshop.

After this workshop, we conducted the second round of interviews (round-2 interviews) with twelve people, five of whom are the same participants from round-1 interviews. This round of interviews was done to assess the feasibility and practical adoption of LLMs in case company 1's RE use cases based on hands-on workshop experiences. It explores necessary tool adaptations, process changes, and lessons learned from using LLMs. They were conducted as four group semi-structured interviews with three participants in each interview. Each interview lasted between 40-50 minutes.

The interviews conducted within case 2 and case 3 explore the use of LLMs in SE practice, focusing on their impact, benefits, and challenges. We aimed at understanding the participants' roles, the LLM-based tools they use, and their specific use cases, workflow changes, best practices, and strategies for optimising LLM usage while addressing limitations. Findings gathered from these interviews helped in refining approaches for adopting LLMs within SE activities. Each interview lasted between 30-40 minutes and was conducted virtually via video conferencing.

\subsection{Data Analysis} \label{subsec:dan}

We conducted reflexive thematic analysis~\citep{Clarke2014} on the data gathered from the interviews to derive findings. This approach treats coding and theme development as interpretive activities carried out by the researcher, rather than as procedures intended to measure agreement or reliability between coders. The interview transcripts were edited, validated through member checking~\citep{799955}, and anonymised to ensure confidentiality. The analysis was first carried out manually by a single author. The first author thoroughly familiarised themselves with the data from an interview transcript to gain an overall sense of the content. During this phase, notes were taken to document emerging observations and analytic decisions to maintain transparency and traceability in the analysis process.

Then, the author began extracting relevant and meaningful segments of text, i.e., the verbatim source quotes from the interview transcript (which we refer to as findings within this study). An example finding is: ``\textit{I still reviewed the code line by line to ensure it does what I expect. I'm still involved in the process and treat the LLM as an assistant, not a decision-maker.}'' This process was repeated for all fifteen interview transcripts from all three cases.

Then, similar source quotes (findings) from all fifteen interviews were collated together under a distinct and meaningful code. For example, the quotes ``\textit{I think AI can't do that (the task) for me, but it can support me in this [task]}'', ``\textit{I could imagine an AI serving as a co-pilot, checking for inconsistencies in diagrams or other output artefacts}'', and ``\textit{In essence, this tool (the LLM) would act like an assistant or co-pilot. It wouldn't automate things}'' were collated together under the code ``Role''. The first two quotes were gathered from two different participants from case 1 and the third quote was gathered from a participant from case 3. Codes were developed inductively by comparing quotes across all interviews and refining code definitions as analysis progressed. This process resulted in a total of 13 distinct and meaningful codes. The code descriptions (code book) are presented in Table~\ref{tab:Table-4}. Table~\ref{tab:Table-5} in subsection~\ref{subsec:agg} provides the traceability of findings to the codes and themes under which they are categorised under. The codebook.xlsx file provided within the online repository\footnote{Zenodo repo with codebook.xlsx file:\url{https://doi.org/10.5281/zenodo.15754264}} provides a trail of the development of themes and codes to the findings along with the source of the findings as well.

\begin{table}[ht!]
\centering
\begin{tabular}{|p{3cm}|p{9.5cm}|}
\toprule
\textbf{Code} & \textbf{Meaning} \\
\midrule
Role & Expectations and perception about the kind of role LLMs should or will play in SE. \\ \hline
Criteria    &  The criteria under which LLMs can employed in SE. \\ \hline
Benefit &  The benefits of employing LLMs in SE. \\ \hline
Method  &  The evaluation methods to assess LLMs' outputs. \\ \hline
Metric  &  The metrics using which LLMs' outputs are evaluated. \\ \hline
Model selection &  Insights on how to select the appropriate LLM based on the SE task at hand. \\ \hline
Implementation  &  Strategies and techniques to effectively use LLMs in SE. \\ \hline
Mandatory requirement   &  Mandatory practices when employing LLMs in SE. \\ \hline
Autonomy    &  The need for human autonomy when employing LLMs in SE. \\ \hline
Changing user responsibilities  & The effect on user responsibilities when employing LLMs in SE. \\ \hline
Changing workflows  &  The effect on development workflows when employing LLMs in SE. \\ \hline
Facilitation    &  How to facilitate training to use LLMs more effectively in SE. \\ \hline
Requirements    &  The necessary skills users need to have to use LLMs effectively in SE.\\
\bottomrule
\end{tabular}
\caption{Thematic analysis codebook.}
\label{tab:Table-4}
\end{table}

To improve transparency and to check that the development of codes and themes were understandable beyond the primary analyst, a second author became involved in the coding process at this stage. Based on the guidelines for assessing and reporting reliability of coding procedures in qualitative research~\cite{10.1111/j.1468-2958.2002.tb00826.x, 10.1145/3359174}, the involvement of the second author was planned for consensus building. The goal was to assess whether the codes and themes could be reasonably interpreted similarly by another researcher and to reduce the risk that the analysis relied on implicit knowledge held only by a single author. No major inconsistencies in the coding and interpretation process were observed between both the researchers. While in some instances both researchers coded the same data extract using different labels, the definitions of these codes were semantically similar. In such cases, the researchers reached a consensus on a single code name, which was then used consistently to code similar data extracts in subsequent analysis as well re-checking the existing annotations to ensure they align with the updated code names and/or definitions. Any such disagreements between the two authors were discussed and resolved.

Once coding was complete, the researchers reviewed and collated codes that address the same aspect into potential themes, considering how different codes combined to capture significant patterns within the data. To reach a consensus in interpretation, we also had regular team discussions and iterative reviews of the developing themes between four of the six involved authors. Finally, the themes were clearly defined and named, with representative quotes from participants selected to illustrate each theme as presented within Table~\ref{tab:Table-5} and Table~\ref{tab:Table-6} in Section~\ref{sec:mcs}. This resulted in the creation of six distinct and exclusive primary themes as follows:

\begin{enumerate}[leftmargin=1.1cm]
\setlength{\itemsep}{0em}
\item The ``\textbf{AI Assistant}'' theme encompasses findings about practitioners' expectations and findings on utilising LLMs as AI assistants within their SE activities rather than automation tools.
\item The ``\textbf{Evaluation}'' theme refers to the methods and metrics for assessing various aspects of the LLM-generated artefacts.
\item The ``\textbf{Applicability}'' theme encapsulates what users expect the capabilities of the LLMs to be and the boundary of the LLMs' applicability for the selected use cases.
\item The ``\textbf{Human Oversight and Agency}'' theme captures findings related to the role of users in overseeing and guiding LLM-generated outputs. It encompasses practitioners' perspectives on maintaining control, ensuring accountability, and balancing AI assistance with human expertise in SE activities. 
\item The ``\textbf{LLM Effect on Workflows}'' theme refers to the impact of LLM integration on existing SE workflows. It includes observations on how LLMs reshape task execution as well as potential adaptations required for efficient LLM adoption.
\item The ``\textbf{User Skills}'' theme encapsulates findings into the skills and knowledge practitioners need to effectively utilise LLMs in SE.
\setlength\itemsep{1em}
\end{enumerate}

Our analysis of the multi-case study data revealed several similar findings across the three cases within each primary theme. Each finding extracted from the interviews that contributed to the synthesis of the recommendations was given an identifier in the format of CXFY. CX, where X = [1,3] refers to the case from which the findings was extracted. FY, where F = [1,22], refers to the specific finding within each case. So, C1F1 represents the first finding from case 1 that was found relevant to the synthesis of a recommendation. Based on these common findings, we synthesised seven recommendations to support the adoption of LLMs in SE activities within industrial settings. Every recommendation is supported by multiple findings from different participants from two or more cases. The seven recommendations derived following this process are presented in Section~\ref{sec:recs}.

\subsection{Validation Survey} \label{subsec:survey}

We conducted an online survey to assess the perceived relevance of the seven recommendations. The survey questionnaire had a total of ten closed-ended questions. The first question, \textbf{Q1}, was a multiple choice single selection question aimed to gather the participant's demographic background, i.e., the industry in which they work. The next two questions, \textbf{Q2} and \textbf{Q3}, were control questions focusing on checking whether the respondents use enterprise versions of the LLMs within their SE use cases and how familiar they are with prompt engineering. Then, in the next seven questions, \textbf{Q4} to \textbf{Q10}, we presented the recommendations and asked the participants to rate their agreement with the seven recommendations on a scale of 1-5, where 1 is ``strongly disagree,'' 2 is ``somewhat disagree,'' 3 is ``neutral,'' 4 is ``somewhat agree,'' and 5 is ``strongly agree.'' 

The target population of the survey were software practitioners (e.g. software developers, requirements engineers, product owners, solution architects, and DevOps engineers) from various industries. A pilot study of this survey was administered to three people to ensure the questions in the survey were understandable and no ambiguity was present. The pilot round of survey was active for one week. Based on the feedback from the pilot respondents, we added examples for the recommendations at end of the questions to demonstrate how they can be applied in industrial practice. Four of the six authors of this study contributed to the development, refinement and validation of the survey questionnaire. Once we were satisfied with the changes made, we distributed the survey online to collect the data. The participants were recruited from our professional network, following a purposive sampling approach~\citep{10.1007/s10664-021-10072-8}, based on fulfilling this criteria: have experience using any LLM-based tool within their work. The survey was also shared on \textit{LinkedIn} with the criteria highlighted in the post description. We followed the ethical interview checklist~\citep{8870192} to collect the participants' informed consent for processing for their responses to synthesise the results. The survey was open for seven weeks.

We used the data collected from this survey to strengthen our arguments surrounding the perceived relevance of the recommendations. The survey questionnaire along with the collected data is provided within the supplementary data within an online repository\footnote{Zenodo repo with survey data:\url{https://doi.org/10.5281/zenodo.15754264}}.

\subsection{Mapping to AI HLEG Principles} \label{subsec:mapping}

The final phase of our study involves mapping the validated recommendations to the seven Trustworthy AI principles proposed by the AI HLEG. This mapping was conducted after the recommendations had been developed and validated. The recommendations themselves were derived inductively from participants' experiences and perspectives.

Although we were aware of the Trustworthy AI principles prior to the study, these principles were deliberately not introduced during the interviews. This decision was made to avoid influencing participants' responses or framing their experiences in terms of predefined regulatory concepts. As a result, participants did not consistently reference or recall these principles, and the resulting recommendations primarily reflect practical concerns raised in the case studies rather than compliance considerations.

The objective of the mapping step is therefore descriptive and reflective: to assess which of the seven principles are supported by the empirically derived recommendations and to identify which principles are not addressed. Hence, we are open to the degree of coverage of all seven principles by design, and also allow for extending the recommendations to proactively address principles that did not emerge from the data.

We discuss the impact of this methodological decision in Section~\ref{sec:tai}.

\section{Multi-case Study Results} \label{sec:mcs}

This section presents the results of the thematic analysis performed on interview data gathered from all fifteen interviews from the three cases. We first present the common findings across all three cases within subsection~\ref{subsec:agg}, organised by each of the six common themes, before presenting case specific findings for each of the three cases. The themes and findings presented in this section are summarised in Table~\ref{tab:Table-5}. The source of the direct quotes supporting the presented findings are denoted with an interview ID: IX, with X ranging from 1 to 15, e.g., I3 refers to the 3rd interview. 

\subsection{Cross Case Analysis} \label{subsec:agg}

[\textbf{AI Assistant}]: Practitioners view LLMs as tools or assistants that enhance their work, providing guidance and suggestions rather than replacing their decision making. As stated in I5, ``\textit{In essence, this tool (the LLM) would act like an assistant or co-pilot. It wouldn't automate things.}'' This highlights the potential for AI to become an interactive and collaborative partner in software development. There is a strong preference for using LLMs in a supervised manner, with concerns arising when they operate without human oversight. Echoing this sentiment, I2 noted, ``\textit{I see no concerns and quite some advantages in using them supervised or in a supporting role.}''

[\textbf{Evaluation}]: The effectiveness of LLMs is judged not only by developers but also by testers, customers, and other stakeholders. As noted in I1, ``\textit{This process (LLMs' output evaluation) would involve not just me, but also expert stakeholders.}'' Rather than demanding flawless outputs, practitioners emphasised whether LLM suggestions are useful and offer valuable insights. Echoing this view, I6 remarked, ``\textit{The [LLM] tool doesn't need to write a perfect user story for me; it just needs to provide me with useful information in an intelligent way.}'' This suggests that LLMs are expected to assist rather than fully automate SE tasks, with usefulness regarded as a more meaningful evaluation criterion than strict accuracy.

[\textbf{Applicability}]: Practitioners recognise that different LLMs are optimised for specific tasks, and their effectiveness varies depending on the domain. A one-size-fits-all approach is not viable, and selecting the right model for a given task is crucial for achieving optimal performance. As noted in I15, ``\textit{Amazon Q developer and GitHub Copilot are designed for different things, so I use each specifically for its intended purpose.}'' Using LLMs beyond their intended scope can lead to suboptimal results. This concern was highlighted in I7, where it was stated, ``\textit{The copilot also expressed dissatisfaction about being misused, so that's probably not a viable approach moving forward.}'' Rather than relying on LLMs for end-to-end solutions, practitioners break down larger goals into smaller steps and use LLMs selectively at each stage. As explained in I15, ``\textit{I break down a bigger goal into smaller steps and tackle them one at a time.}'' This reflects a structured and controlled approach to leveraging LLM assistance.

[\textbf{Human Oversight and Agency}]: Practitioners stress the importance of human involvement in reviewing and verifying LLM-generated outputs. Despite LLMs' impressive capabilities, human judgment remains critical, especially in large-scale or complex applications. As emphasised in I14, ``\textit{There's still a need for human oversight and verification, especially when it's being used on a large scale and by people who may not fully understand how it works.}''

[\textbf{LLM Effect on Workflows}]: The adoption of LLMs is expected to cause significant changes in existing workflows. Organisations need to carefully evaluate and adapt their current processes to incorporate LLMs effectively, which may require restructuring or redesigning workflows to accommodate LLMs' capabilities. As noted in I8, ``\textit{We need to carve out a space for LLMs in our business process \dots our existing processes would need to be restructured.}'' There is uncertainty around how the new LLM-assisted workflows should be designed and developed. This challenge was highlighted in I11: ``\textit{Another challenge is that it changes the workflows, and agreeing on what the new workflows should be is tough.}'' As LLMs advance, the role of developers is expected to shift. According to I3, ``\textit{If we have less work with basic or standard activities, we could focus more on our main task, which is to talk to our customers.}'' Developers may no longer need to engage in the more repetitive groundwork, instead focusing on validating and ensuring the quality of LLM-generated outputs.

[\textbf{User Skills}]: Practitioners emphasise the need for proper training on how to interact with LLMs. As communication with these tools differs significantly from human interactions, learning how to engage with them effectively is critical for maximising their value. As stated in I8, ``\textit{Provide proper education about them. If you don't know how to use it, in most cases, it will be completely useless.}'' A strong understanding of the domain is considered essential for using LLMs effectively. This was highlighted in I15: ``\textit{LLMs aren't just plug-and-play; you need to understand how they work, how to prompt them correctly, and where they can actually add value.}'' Practitioners also stress the importance of sharing insights on what works well and what does not. As noted in I15, ``\textit{We often share ideas on what works well for specific use cases.}'' Effective use of LLMs involves not just technical know-how but also collaboration and knowledge sharing within teams to understand where and how AI can add the most value.

\begin{table}[H]
\centering
\resizebox{\linewidth}{!}{
\begin{tabular}{|p{3cm}|p{4cm}|p{4cm}|p{2.5cm}|}
\hline
\multirow{1}{*}{\textbf{Theme}} &
  \multicolumn{1}{c|}{\multirow{1}{*}{\textbf{Code}}} &
  \multicolumn{1}{c|}{\multirow{1}{*}{\textbf{Findings}}} &
  \multicolumn{1}{c|}{\multirow{1}{*}{\textbf{Source Case}}} \\ \hline
\multirow{3}{*}{AI Assistant}            &  Role &  C1F1- I5; C1F2 - I3; C1F3 - I2, C2F1- I12;  C3F1- I15 & Case 1, Case 2, Case 3                     \\ \cline{2-4}  
                                         &  Criteria     &       C1F4 - I2; C1F5 - I6; C3F2 - I14       & Case 2, Case 3         \\ \cline{2-4} 
                                         &  Benefit   &   C1F6 - I7; C2F2 - I12; C3F3 - I14; C3F4 - I15    & Case 1, Case 2, Case 3                           \\ \hline
\multirow{2}{*}{Evaluation}              & Method  & C1F7 - I1; C1F8 - I4; C1F9 - I5; C3F5 - I15  &  Case 1, Case 3 \\ \cline{2-4}
                                        & Metrics & C1F10 - I6; C1F11- I5; C2F3 - I11 & Case 1, Case 2 \\ \hline
\multirow{2}{*}{Applicability}          & Model Selection & C1F12- I7; C1F13- I8; C3F6 - I15  & Case 1, Case 3 \\ \cline{2-4} 
                                         & Implementation & C2F4- I 12; C3F7 - I15; C3F8 - I15                        & Case 2, Case 3                    \\ \hline
\multirow{2}{*}{\parbox{6cm}{Human Oversight \\ and Agency}}        &  Mandatory Requirement & C1F14- I1; C1F15- I4; C2F5- I12; C2F6- I12; C3F9- I14; C3F10- I14 & Case 1, Case 2, Case 3            \\ \cline{2-4}
                                                    & Autonomy &    C1F16 - I1     & Case 1     \\ \hline
\multirow{2}{*}{\parbox{6cm}{LLM Effect on \\ Workflows}} & Changing user responsibilities & C1F20- I8; C2F8 - I12 & Case 1, Case 2 \\ \cline{2-4}
                                        & Changing Workflows & C1F17-I8; C1F18- I8; C1F19- I9; C2F7- I11; C3F11- I15; C3F12- I14; C3F13 - I14  & Case 1, Case 3 \\\hline
\multirow{2}{*}{User skills}           & Facilitation & C1F21- I9; C1F22- I8; C3F14- I14; C3F15- I14; C3F16- I15 & Case 1, , Case 2, Case 3    \\ \cline{2-4} 
                                         & Requirements & C2F9- I12; C2F10- I13; C3F17- I14; C3F18- I15  & Case 2, Case 3  \\ \hline
\end{tabular}
}
\caption{Primary themes, the corresponding codes and findings, and the source(s) from which the findings were extracted.} \label{tab:Table-5}
\end{table}

\subsection{Case 1}

The findings specific to case 1 are based on interviewees' understanding of LLMs, prompt engineering, and RE artefacts, tasks, and processes at the case company before and after testing LLM-assisted RE use cases in the workshop. These findings focus on (i) expected improvements in their RE process with LLM assistance, (ii) planned evaluation methods for LLMs' outputs, and (iii) qualities of LLMs prioritised for real-world adoption. 

Practitioners believe LLMs can perform analogical reasoning tasks, such as generating high-level project requirements and deriving insights from past projects. They also see potential in using LLMs to identify redundancies in requirements elicitation meetings and system requirements specifications (SRS). Multiple participants expressed interest in using LLMs to check the completeness of RE artefacts like SRS. As stated in I5, ``\textit{This tool could ask us questions or we could ask it questions, which could help us remember ideas or aspects of our ecosystem that we might have overlooked. It could also help us learn from past projects, both successful and unsuccessful.}'' After using three LLM-based tools during the workshop, participants believed LLMs could aid RE processes by retrieving and analysing data, improving productivity, providing different perspectives, and enhancing coverage of overlooked aspects.

Participants also showed interest in automating tasks such as generating artefacts (e.g., user stories) and checking artefact consistency and dependencies for requirements traceability and change management. As noted in I4, ``\textit{When it comes to requirements, having support in creating them can be very helpful. If we do not have to worry about the formal structure, or if the acceptance criteria and affected systems are correctly mentioned, it can save a lot of time.}''

The RAG-POC tool required more detailed input prompts than Microsoft Copilot to convey users' intentions and sometimes failed to provide answers consistently for the same prompts. Participants attributed these challenges to the lack of sufficient stakeholder analysis artefact data within the RAG database and felt the tool needs greater access to company-specific data and knowledge. This was reflected in I6: ``\textit{While the user story use case worked well, the generating stakeholder analysis artefacts use case was a bit more complicated, and we lacked sufficient knowledge about the information already provided and what we could upload [to the RAG database].}'' This highlights the challenges faced with the RAG POC tool.

Efficiently setting up the RAG-POC tool and modifying the RAG architecture to include company-specific additions can be a challenging task due to its complexity. The construction and curation of the knowledge base to be used for RAG is currently presenting research challenges~\citep{10.1007/978-3-031-49601-1_2}. The findings of this case strengthen the arguments focusing on the lack of methodology that guides organisations on how to identify valuable and relevant information for developing useful RAG architectures.

\subsection{Case 2}

In this case, LLMs were widely valued for their role in learning and cognitive support. Practitioners used them to explore new domains, summarise information, and offload mentally taxing tasks. As noted in I13, ``\textit{For me, the main motivation is to learn new things, summarise information, and explore new fields.}'' LLMs were also observed to reduce cognitive strain on users. According to I10, ``\textit{It eases the cognitive load because I don't have to remember all the command lines when it comes to coding.}''

While LLMs assist in ideation, maintaining control over their creative outputs remains a challenge. As stated in I11, ``\textit{The LLMs tend to be too creative, and simple prompting isn't enough to make them adhere to these constraints.}'' However, some saw value in leveraging LLMs for creative exploration: ``\textit{I think it would be useful to at least try to perform the [creative] task with the LLM and see if the answer is satisfactory.}'' as expressed in I10. This approach, however, requires strong domain knowledge on the user's part, along with human oversight mechanisms to be advantageous.

Fragmented LLM adoption across teams led to concerns about inefficiency. One participant suggested a more centralised approach: ``\textit{Many different teams want to try it out, which leads to multiple LLMs being used across the organisation. Perhaps we shouldn't have so many and instead focus on having one base that can be applied in different contexts.}'' as stated in I11. Others emphasised testing high-value use cases before scaling: ``\textit{Instead of evaluating all possible use cases at once, we should focus on one small, high-value use case.}'' also from I11. In contrast, some practitioners experimented with LLMs interacting as specialised agents to refine results. One described their approach: ``\textit{I'm using one LLM to optimise the input for another. It's like an agent approach, where each tool works together to improve the end result.}'' as mentioned in I12. Another proposed developing domain-specific agents: ``\textit{We need one agent for generating standard documentation and another for troubleshooting content.}'' from I11. Overall, while diversification of LLM models across organisations adds complexity to the adoption process, utilising multiple LLMs in an agentic approach or similar can contribute to the resilience of LLM-assisted workflows by mitigating the limitations that come with relying on a single LLM model.

The value LLMs brought to SE was seen to be assessed through multiple factors, including efficiency, cost, and business impact. One team developed a structured approach: ``\textit{We look at value from three angles: what it brings to the customer, what it brings to the company in terms of profitability, and the trade-offs in development and maintenance costs.}'' as stated in I11.

LLMs were also observed to impact the user's psychological aspects such as increasing confidence when tackling new or complex tasks. As reflected in I13, ``\textit{I feel more confident taking on new or unknown tasks}'' and ``\textit{It also helps with confidence since I have an initial helper before reaching out to colleagues.}''

\subsection{Case 3}

Similar to case 2, practitioners from case 3 also found LLMs particularly useful for quickly grasping new topics without sifting through extensive online resources. As noted in I15, ``\textit{If I don't know anything about it (a new task), it's easier to go to the LLMs instead of googling and reading through all the stuff.}'' Additionally, the effect of LLMs on cognitive load was similarly reflected: ``\textit{I feel a lot less overwhelmed when dealing with unfamiliar tasks.}'' as expressed in I15.

However, unlike case 2 where most practitioners preferred the auto complete feature of Copilot for coding assistance, many users from case 3 favoured chat-based interactions over code completion due to greater context awareness. According to I15, ``\textit{I use the chat option within Copilot a lot. I don't use the code completion much because it lacks the context awareness that the chat option provides. For me, the multiple suggestion chat is more useful than the code completion.}''

LLMs were recognised for improving efficiency and reducing workload. As stated in I14, ``\textit{I think LLMs can really help cut down on day-to-day work and make things more efficient.}'' However, limitations in contextual understanding were also highlighted: ``\textit{The AI doesn't think the way a developer does. It doesn't have context or real understanding of the system.}'' as noted in I14.

Training provided to users regarding LLM usage was limited more to compliance than practical application. As mentioned in I15, ``\textit{We received some training on using AI, but it was mostly focused on what not to do, along with legal guidelines.}'' Additionally, the abundance of LLM tools available in the market was observed to create confusion around adoption strategy. This concern was expressed in I14: ``\textit{With so many models out there, it can be a bit confusing to know which one to choose.}''

\begin{summarybox}
\textbf{Summary of RQ1 results:}  
The results of our multi-case study highlight that practitioners prefer to use LLMs as an assistant, prioritise the usefulness of the outputs over their accuracy, follow a structured approach to leveraging LLMs within a defined scope of applicability, focus on understanding and adapting to effects of LLMs on traditional SE workflows, emphasise the need to develop human oversight mechanisms and the necessary skills to leverage LLMs effectively.
\end{summarybox}

\section{Recommendations from the Multi-Case Study} \label{sec:recs}

From the themes and findings, we were able to derive seven recommendations. Each recommendation is mapped to one the themes from our case analysis. Some themes were complex enough to warrant more than one recommendation. Each recommendation is supported by findings from at least two cases. The recommendations are as follows:

\begin{enumerate}[leftmargin=1.1cm]
\setlength{\itemsep}{0em}
 \item [\textbf{R1}:] It is beneficial to use the LLMs as assistants with human oversight rather than try and automate the task using the LLMs.
 \item [\textbf{R2}:] When using LLMs as AI assistants, it is better to prioritise the usefulness of the LLMs' outputs rather than output accuracy.
 \item [\textbf{R3}:] The relevant stakeholders' satisfaction should be used as part of the LLM-generated outputs evaluation criteria.
 \item [\textbf{R4}:] It is not beneficial to use any task-specific fine-tuned LLMs like GitHub Copilot for tasks they were not designed for.
 \item [\textbf{R5}:] It is crucial to develop and implement human oversight and validation mechanisms while using LLMs' assistance for any task.
 \item [\textbf{R6}:] There is a need to create room for incorporating LLMs into business processes by restructuring (parts of) them.
 \item [\textbf{R7}:] Knowledge-sharing activities like seminars and specialised training are crucial for the users of the LLMs.
\setlength\itemsep{1em}
\end{enumerate}

\normalsize

The mapping of the recommendations to the six primary themes and supporting quotes can be found in Table~\ref{tab:Table-6}.

\footnotesize

\begin{longtable}{|p{.2\textwidth}|l|p{.7\textwidth}|}
\hline
\textbf{Theme} & \textbf{ID} & \textbf{Supporting quotes} \\ \hline
\endhead
\hline

\caption{Supporting quotes from one or more of the case companies for each recommendation (continues in the next page).} \label{tab:Table-6}
\endfoot
\hline
\caption{Supporting quotes from one or more of the case companies for each recommendation.} \label{tab:Table-6}
\endlastfoot

\textbf{AI Assistant} & \textbf{R1} & \textbf{C1F1}: ``In essence, this tool (the LLM) would act like an assistant or co-pilot. It wouldn't automate things'' \newline \textbf{C1F2}: ``I think AI can't do that (the task) for me, but it can support me in this [task]'' \newline \textbf{C2F1}: ``I see LLMs as a helper tool'' \newline \textbf{C3F1}: ``I would love to have something like a coding assistant.'' \\ \hline
\textbf{Evaluation} & \textbf{R2} & \textbf{C1F7}: ``This process (LLM's outputs evaluation) would involve not just me, but also expert stakeholders.'' \newline \textbf{C1F8}: ``ask software testers, not developers, if the solution meets the requirement. Testers usually have a better understanding of how things are connected and what's needed to reach the goal.'' \newline \textbf{C1F9}: ``I plan to conduct feedback (for generated output) sessions with different relevant stakeholders to determine its quality and usefulness'' \newline  \textbf{C3F5}: ``I always prefer a domain expert to review it [LLM's output] during code review.''\\ \hline
\textbf{Evaluation} & \textbf{R3} & \textbf{C1F10}: ``The tool doesn't need to write a perfect user story for me; it just needs to provide me with useful information in an intelligent way.'' \newline \textbf{C1F11:} ``the tool should be helpful to those who use it'' \newline \textbf{C2F3}: ``we focus on desirability: does it help end users?'' \\ \hline
\textbf{Applicability} & \textbf{R4} & \textbf{C1F12}: ``It definitely makes sense to have models specialised for specific tasks in terms of performance'' \newline \textbf{C1F13}: ``GitHub Copilot wasn't the right tool for what we tried because it's more code-based'' \newline \textbf{C3F6}: ``Amazon Q, it's better suited for AWS CDK related tasks. Amazon Q helps specifically with the Amazon-related things, because it's trained on that.'' \newline \textbf{C3F7}: ``Amazon Q and GitHub Copilot are designed for different things, so I use each specifically for its intended purpose.'' \\ \hline
\textbf{Human Oversight and Agency} & \textbf{R5} & \textbf{C1F14}: ``there's still a need for human oversight and verification, especially when it's being used on a large scale and by people who may not fully understand how it works'' \newline \textbf{C1F15}: ``But ultimately, human feedback is crucial.'' \newline \textbf{C2F5}: ``I'm definitely in favor of keeping humans in the loop.'' \newline \textbf{C2F6}: ``I still reviewed the code line by line to ensure it does what I expect. I'm still involved in the process and treat the LLM as an assistant, not a decision-maker.'' \newline \textbf{C3F9}: ``end of the day, you're still responsible for checking and making sure everything works as expected.'' \newline \textbf{C3F10}: ``Whatever code an LLM generates, you are still responsible for it.'' \\ \hline
\textbf{LLM Effect on Workflow} & \textbf{R6} & \textbf{C1F17}: ``our existing processes would need to be restructured'' \newline \textbf{C1F18}: ``We need to carve out a space for LLMs in our business process'' \newline \textbf{C1F19}: ``The integration of such tools into our workflow would likely require a careful evaluation and adjustment of our current processes.'' \newline \textbf{C2F7}: ``Another challenge is that it changes the workflows, and agreeing on what the new workflows should be is tough.'' \newline \textbf{C2F8}: ``As LLMs become more advanced, I think we'll see a shift where developers to review the code and ensure things are working properly, but they won't be doing the groundwork'' \newline \textbf{C3F11}: ``AI is influencing our planning process as well. We're seeing a shift in traditional practices due to LLMs, even in the planning stage.'' \newline \textbf{C3F12}: ``we may not yet be fully considering how it will impact our workflows and which use cases will provide the most benefit.'' \newline \textbf{C3F13}: ``We're still figuring out how to design software that's LLM-friendly. If we were to build something new today, how do we ensure it can be integrated with an LLM down the line? That's a question we don't fully have the answer to yet.'' \\ \hline
\textbf{User Skills} & \textbf{R7} & \textbf{C1F21}: ``there's a need for training on how to interact with these tools, as communicating with them is not the same as conversing with another person'' \newline \textbf{C1F22}: ``provide proper education about them. If you don't know how to use it, in most cases, it will be completely useless'' \newline \textbf{C2F9}: ``I think the key is to understand what you're doing. LLMs can generate code, but it may become too complex for you to fully comprehend. If that happens, you're probably already in over your head.'' \newline \textbf{C2F10}: ``I think the main thing is to have domain knowledge before using LLMs.'' \newline \textbf{C3F14}: ``one of the biggest challenges is getting people up to speed on how to use LLMs effectively. It takes proper planning, understanding, and the right implementation to really make use of it.'' \newline \textbf{C3F15}: ``Part of our responsibility is to stay updated on these tools and understand how they can help us. We're even getting some training on generative AI to see how it could fit into our work.'' \\
\textbf{User Skills} & \textbf{R7} & \textbf{C3F16}: ``we often share ideas on what works well for specific use cases'' \newline \textbf{C3F17}: ``knowing which tool to use and understanding what a model is trained on are key'' \newline \textbf{C3F18}: ``LLMs aren't just plug-and-play; you need to understand how they work, how to prompt them correctly, and where they can actually add value.'' \\ \hline
\end{longtable}

\normalsize

\begin{summarybox}
\textbf{Summary of RQ2 results:}  
We have synthesised seven recommendations to support efficient and responsible adoption of LLMs within industrial software development. These recommendations are aimed to guide practitioners when deciding how to use LLMs, determining the scope of LLM's application and criteria for evaluating LLMs' outputs, emphasising necessary considerations to enhance regulatory compliance, ease LLM integration into business processes, maximise the adoption benefits, and improve overall user experience and ease of use.
\end{summarybox}

\section{Survey Results} \label{sec:survey}

This section presents the results of the online survey conducted to understand the extent to which the recommendations from the case study are applicable to contexts outside of the selected case companies. Subsection~\ref{subsec:demo} presents the demographics of the survey participants. Subsection~\ref{subsec:insights} presents the quantitative results of the survey responses.

\subsection{Participants' Demographic Information} \label{subsec:demo}

\begin{figure*}[ht!]
 \centering
\includegraphics[width=\linewidth,trim={0 0 0 1.2cm},clip]{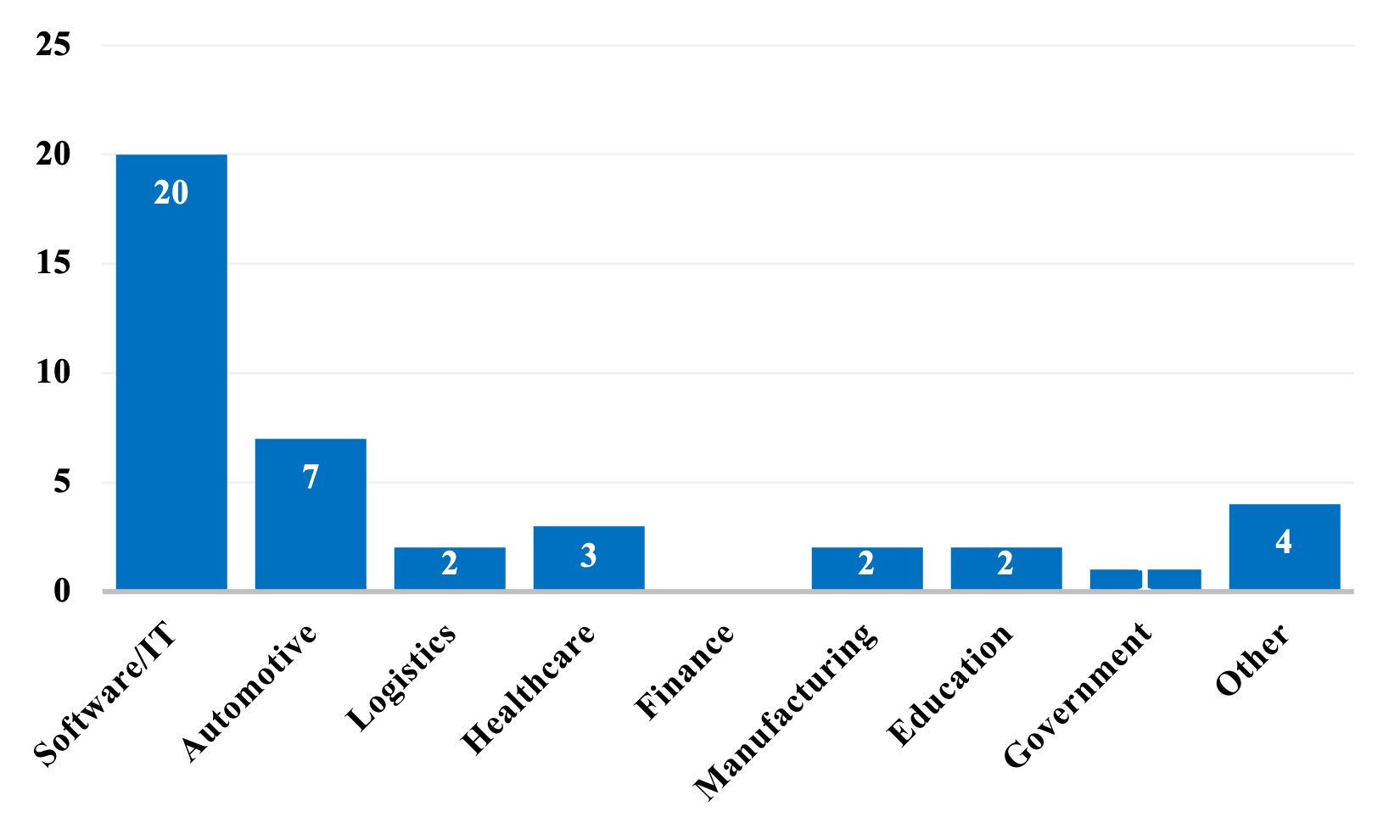}
\caption{Distribution of respondents based on the industry they are currently working in.} \label{fig:fig-2}  
\end{figure*}

\begin{figure*}[ht!]
 \centering
\includegraphics[width=\linewidth,trim={0 0 0 1cm},clip]{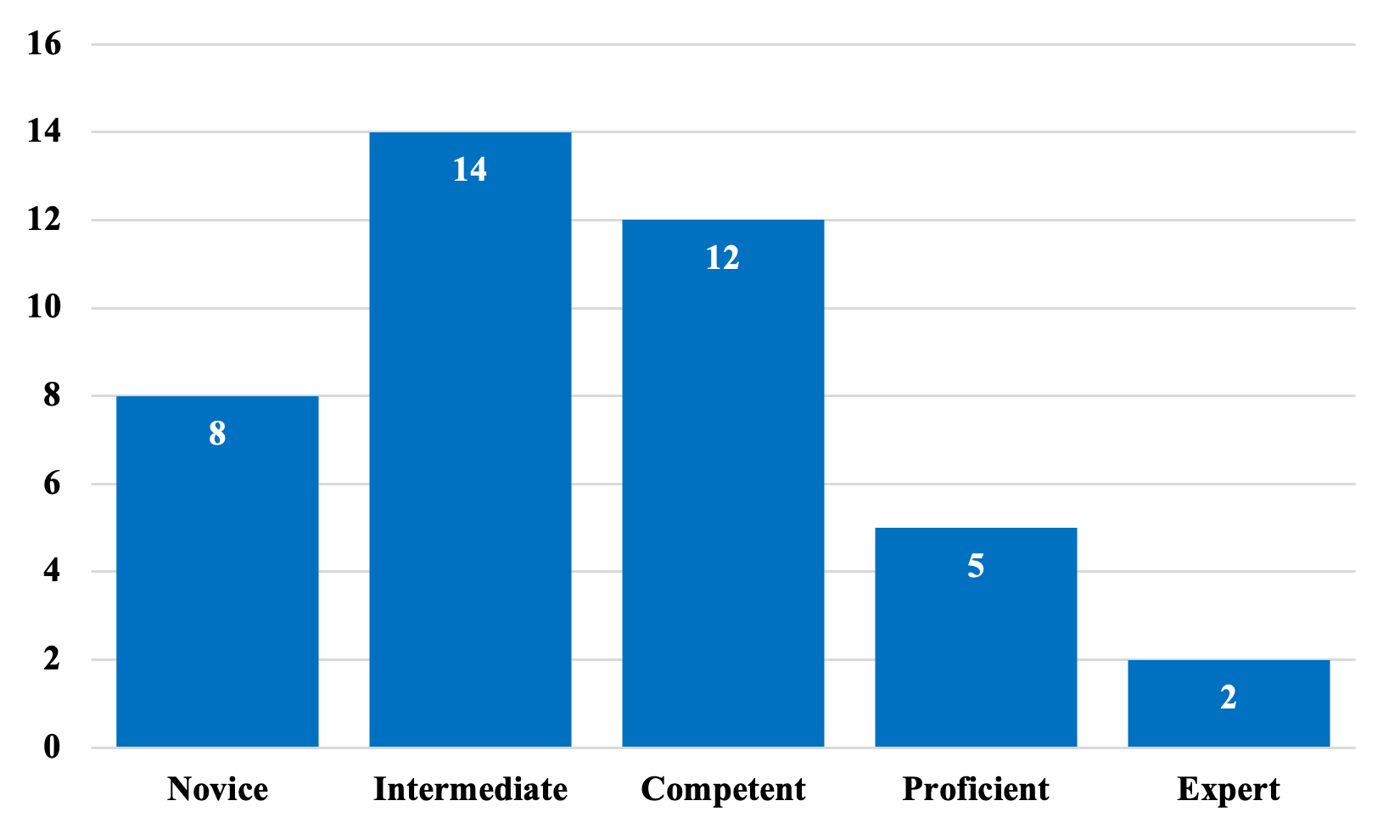}
\caption{Distribution of respondents based on their self identified level of skill within prompt engineering.} \label{fig:fig-3}  
\end{figure*}

We collected a total of 43 responses over a span of seven weeks, out of which 41 responses were usable; we discarded two responses because they were incomplete. The median response time for answering the survey was 5 minutes 31 seconds. The complete breakdown of the demographic data for the survey participants is presented within Figure~\ref{fig:fig-2} and Figure~\ref{fig:fig-3}. 

Figure~\ref{fig:fig-2} represents the demographic breakdown of survey participants by industry. Out of the 41 respondents, 20 indicated working in the Software/IT services sector, which we use to refer to organisations providing IT services or consultancy not dedicated to a specific application domain (e.g. Automotive, Finance, Healthcare). The second highest number of respondents came from the Automotive industry with 7 participants. The Manufacturing and Education sectors each had 3 participants, accounting for 7.3\% of responses, while the Logistics, Healthcare, and Government sectors each had 2 participants, representing 4.87\% of responses. Notably, although Finance was provided as an option, no respondents selected it. Lastly, the category labelled `Other' includes 4 participants who indicated domains outside the predefined options. This distribution shows a significant inclination of survey responses from individuals in the Software/IT sector compared to other sectors. 68\% of participants reported to have used or using an enterprise version of LLM-based tools such as Microsoft Copilot, Github Copilot, or ChatGPT enterprise. Out of the participants who use an enterprise version of the LLM, 70\% of them belong to the software/IT industry.

Figure~\ref{fig:fig-3} represents the self-reported prompt engineering skills of survey participants. The participants rated their skills on a Likert scale from 1 to 5, with 1 being ``Novice'' and 5 being ``Expert''. The majority of participants rated themselves as ``Intermediate'' with a total of fourteen respondents, accounting for 34\% of the total participants. This was closely followed by twelve participants who rated themselves as ``Competent'', making up 29\% of the total participants. ``Novice'' and ``Proficient'' categories had eight and five participants, making up 20\% and 12\% of the total participants respectively. Only two participants rated themselves as ``Expert''. This distribution indicates a moderate level of prompt engineering skills among the survey respondents, with most participants identifying themselves as having intermediate to competent skills.

\subsection{Recommendations' Evaluation} \label{subsec:insights}

\begin{figure*}[ht!]
 \centering
\includegraphics[width=0.9\linewidth]{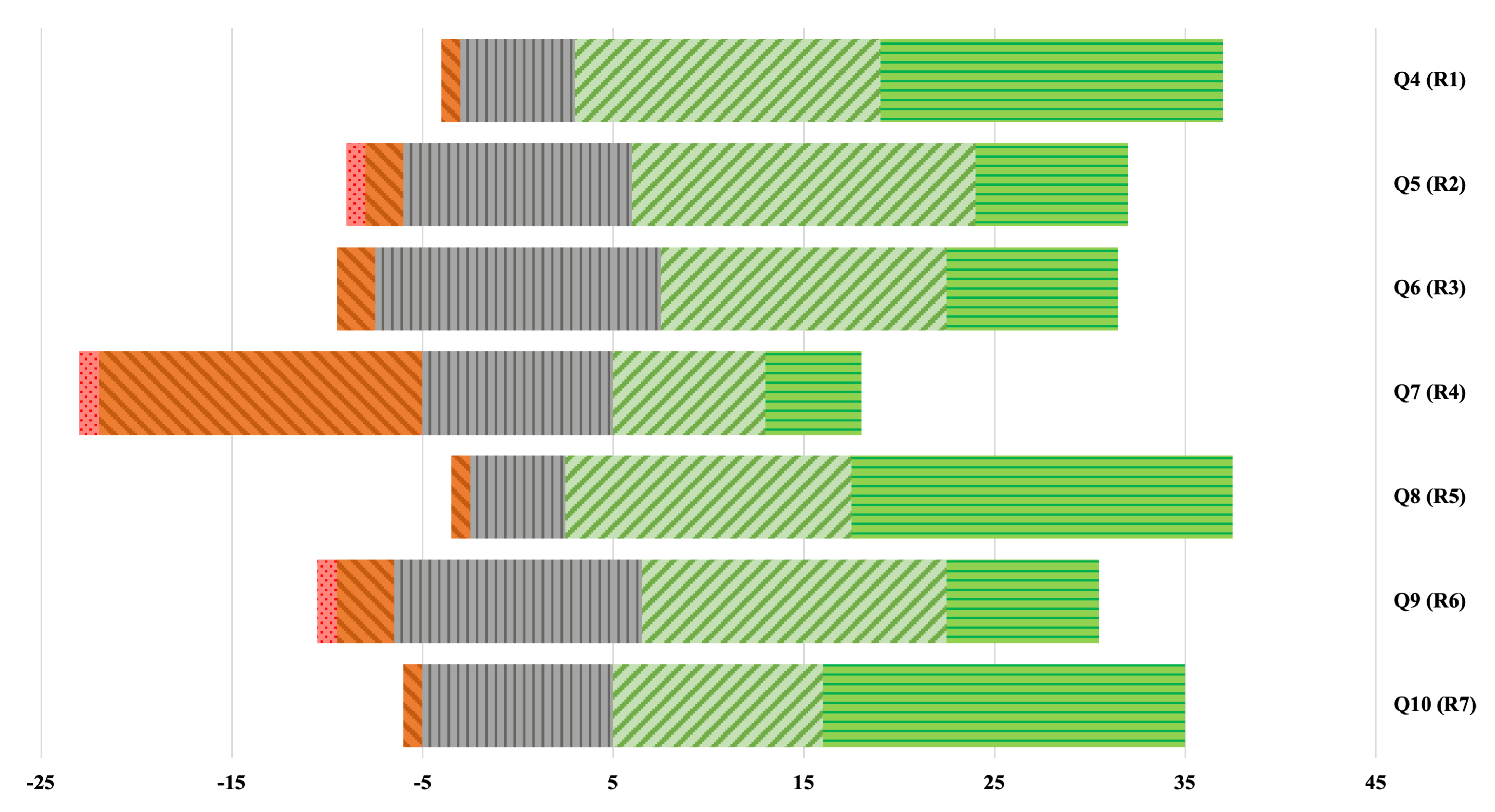}
\caption{Aggregated Likert scale responses for Q7 to Q14} \label{fig:fig-4}  
\end{figure*}

The online survey was meant to explore whether software practitioners outside the case companies also saw value in the recommendations. This section reports the level of agreement of the surveyed practitioners with each of the recommendations \textbf{R1}-\textbf{R7} presented to them within questions \textbf{Q4}-\textbf{Q10} of the survey. Figure~\ref{fig:fig-4} presents the aggregated Likert scale responses from all 41 participants.

Figure~\ref{fig:fig-4} illustrates the distribution of responses for the remaining survey questions, \textbf{Q4} to \textbf{Q10}. The dark green horizontal lines indicate a score of 5, light green upward diagonal lines indicate a score of 4, the grey vertical lines indicate a score of 3, the orange downward diagonal lines indicate a score of 2 and the red dotted pattern indicate a score of 1.

Figure~\ref{fig:fig-4} shows that 18 out of 41 respondents (45\%) for \textbf{Q4} strongly agree with the presented recommendation (survey value of 5), 16 out of 41 respondents (38\%) showed moderate agreement (survey value for 4), while 6 (15\%) remained neutral (survey value of 3). Only 1 respondent (3\%) somewhat disagreed (survey value of 2) with the recommendation. If we aggregate both strong and moderate agreement (a response of 4 or 5), \textbf{83\%} of respondents for \textbf{Q4} agree with the recommendation. Looking at the demographic details of the participants who gave favourable responses (a response of 4 or 5), 16 out of 34 participants (47\%) are from the software/IT industry.

Similarly, for \textbf{Q5}, 8 out of 41 respondents (20\%) strongly agree, 18 respondents (44\%) show moderate agreement while 12 (29\%) remained neutral. 2 (5\%) somewhat disagree while 1 (2\%) strongly disagree (survey value of 1) with the recommendation. \textbf{64\%} of the total respondents agree with the recommendation. 16 of the 26 (62\%) participants who gave a favourable response are from the software/IT industry. 

For \textbf{Q6}, 9 out of 41 respondents (22\%) strongly agree, 18 respondents (37\%) show both moderate and neutral agreements. Only 2 respondents (4\%) somewhat disagreed with the recommendation. \textbf{59\%} of the total respondents agree with the recommendation. 10 out of the 24 (42\%) participants who gave a favourable response are from the software/IT industry.

For \textbf{Q7}, only 5 out of 41 respondents (12\%) strongly agree, 8 respondents (20\%) moderate agreement while 10 (24\%) remained neutral. 17 respondents (42\%) somewhat disagreed with the recommendation and only 1 (2\%) strongly disagreed. Only \textbf{32\%} of total respondents agree with the recommendation while \textbf{44\%} of them disagreed. 9 out of the 13 (69\%) participants who gave a favourable response are from the software/IT industry.

For \textbf{Q8}, 20 out of 41 respondents (50\%) strongly agree, 14 respondents (35\%) show moderate agreement while 5 (13\%) remained neutral. Only 1 respondent (2\%) somewhat disagreed with the recommendation. \textbf{85\%} of the total respondents agree with the recommendation. 18 out of 35 (52\%) participants who gave a favourable response are from the software/IT industry.

For \textbf{Q9}, 8 out of 41 respondents (20\%) strongly agree, 16 respondents (39\%) show moderate agreement while 13 (32\%) remained neutral. Only 3 respondents (7\%) somewhat disagree and 1 respondent (2\%) with the recommendation. \textbf{59\%} of the total respondents agree with the recommendation. 14 out of 24 participants (58\%) who gave a favourable response are from the software/IT industry.

For \textbf{Q10}, 19 out of 41 respondents (46\%) strongly agree, 11 respondents (27\%) show moderate agreement while 10 (25\%) remained neutral. Only 1 respondent (2\%) somewhat disagreed with the recommendation. \textbf{73\%} of the total respondents agree with the recommendation. 16 out of 30 participants (53\%) who gave a favourable response are from the software/IT industry.

In addition, we conducted subgroup analyses to explore potential differences in perceptions across participants with varying backgrounds. The analysis examined respondents' agreement with the seven recommendations based on three factors: (1) industry (\textbf{Q1}), (2) prompting proficiency (\textbf{Q2}), and (3) access to enterprise LLMs and RAG enhanced LLM system (\textbf{Q3}). For analytical clarity, responses for \textbf{Q2} rated 1–3 (novice to competent) were classified under less proficient users, and ratings of 4–5 (proficient to expert) as highly proficient users in the subgroup analysis.

We applied a one-way ANOVA to assess whether differences in these factors influenced respondents' levels of agreement with the recommendations, followed by Welch's t-tests for pairwise and population-level comparisons. Across the three grouping factors, the results revealed no statistically significant differences in agreement levels, except for two cases associated with prompting proficiency. Participants with higher prompting proficiency showed significantly greater agreement with \textbf{R2} (p = 0.0319) and with \textbf{R3} (p = 0.0169). These findings suggest that individuals with higher prompting proficiency agree more with our recommendations concerning the evaluative aspects of LLM-assisted workflows (\textbf{R2} and \textbf{R3}). This preference might be a consequence of their significant experiences prompting and collaborating with LLMs within SE.

Overall, the strongest agreement was concentrated around recommendations that frame LLMs as assistive rather than autonomous technologies, emphasise human oversight and validation, and encourage knowledge-sharing and training for users. Taken together, this pattern suggests that practitioners see successful LLM adoption in SE less as a matter of simply deploying the technology and more as a matter of integrating it within appropriate human-centric and organisational mechanisms. This is an important finding because it indicates that practitioners associate effective use not only with the technical performance of LLMs, but also with governance, learning, and responsible integration into everyday work.

These findings address RQ3 by showing that the perceived relevance of the recommendations is highest when they address the practical conditions that make LLM-assisted work usable and trustworthy in organisational settings. The results therefore suggest that practitioners place particular importance on recommendations that preserve accountability, support informed human judgement, and help teams develop the skills needed to work effectively with LLMs. At the same time, the subgroup analysis indicates that these perceptions were largely consistent across different respondent backgrounds, which strengthens the view that these concerns are broadly shared rather than limited to a particular subset of participants.

\begin{summarybox}
\textbf{Summary of RQ3 results:}  
 The survey results indicate that most of the recommendations synthesised from the case study are applicable beyond the context of the case company, with six out of the seven recommendations receiving a majority ($>$50\%) agreement from the survey participants. The software/IT industry practitioners make up roughly 48\% of the total sampled population, yet they provided ~54\% of the favourable responses for seven recommendations. Finally, software practitioners with higher prompting proficiency tend to agree more with our recommendations about LLMs' output evaluation more than others.
\end{summarybox}

\section{Trustworthy AI Compliance} \label{sec:tai}

To use LLMs for SE activities in practical settings beyond sandbox environments, software practitioners are recommended to comply with the seven key principles prescribed by the AI HLEG. However, while our findings provide insights relevant to selected principles (e.g., \textbf{R1} and \textbf{R5} emphasise human oversight), the interview data reveal a substantial gap between the AI HLEG principles and the day-to-day concerns and practices of software practitioners.

Motivated by this observation, we examine how the seven empirically derived recommendations relate to the Trustworthy AI principles. Specifically, we performed a post hoc mapping of the recommendations to the seven principles with the goal of identifying which principles are supported by the recommendations and, equally importantly, which are not. This mapping is intended to be descriptive rather than prescriptive and serves to highlight gaps between experience-based guidance and policy-level expectations.

\begin{figure*}[ht!]
 \centering
 \includegraphics[width=\linewidth]{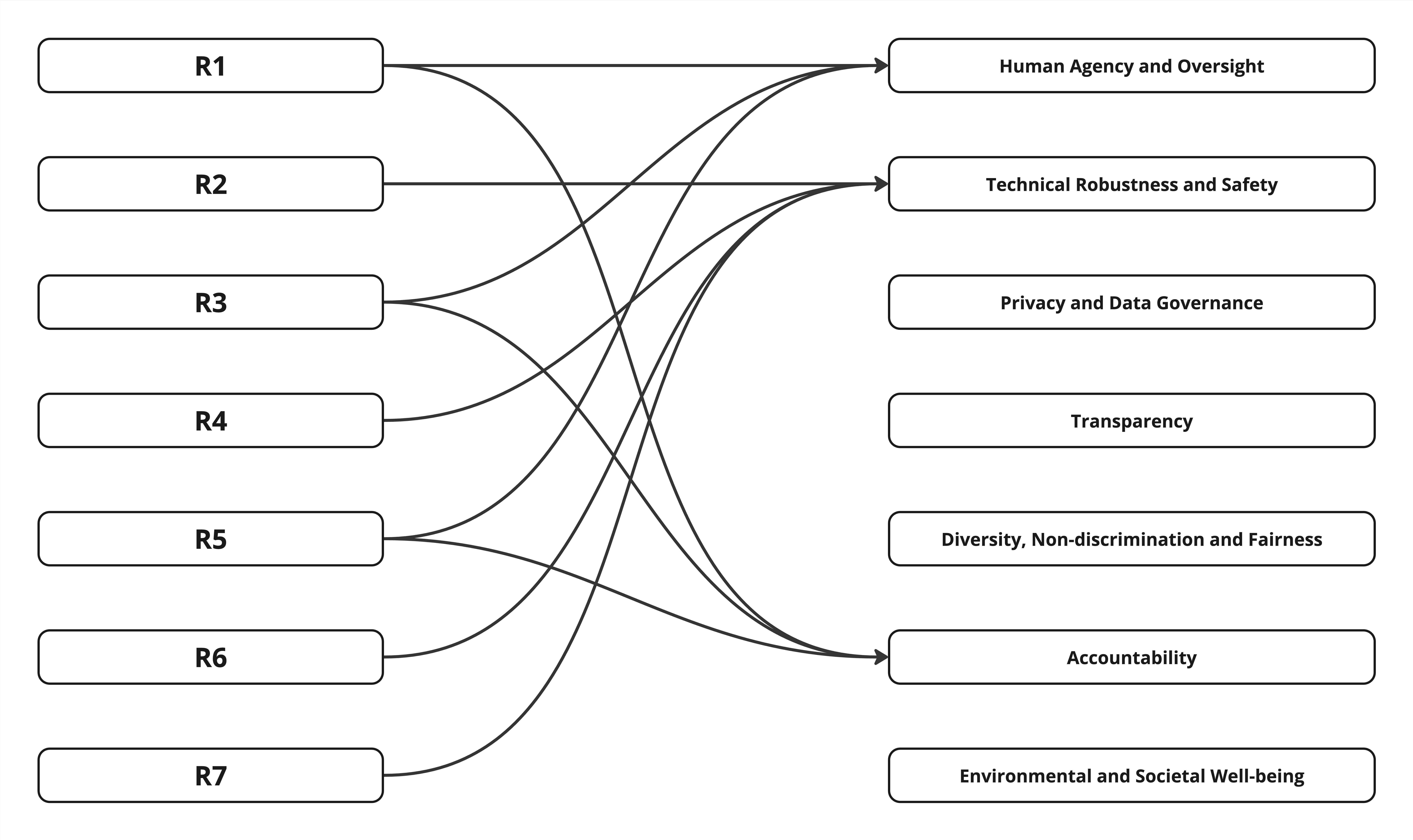}
 \caption{Mapping of the seven recommendations to trustworthy AI principles.}
 \label{fig:fig-5}
\end{figure*}

Figure~\ref{fig:fig-5} illustrates the relationship between the recommendations and the AI HLEG principles. For example, \textbf{R1} and \textbf{R5} emphasise the need for human oversight when using LLMs in SE activities. These recommendations contribute to the principles of \textit{Human Agency and Oversight} and \textit{Accountability}, as responsibility for decisions and outputs remains with the human actor. Similarly, \textbf{R3}, which recommends involving relevant stakeholders in the evaluation of LLM outputs, also aligns with these two principles by reinforcing oversight and shared responsibility.

\textbf{R2}, which guides practitioners on appropriate reliance on LLMs, contributes to the principle of \textit{Technical Robustness and Safety}. Likewise, \textbf{R4}, which limits the use of fine-tuned LLMs to their intended contexts, supports output reliability and thus also aligns with \textit{Technical Robustness and Safety}. \textbf{R6} highlights the need to restructure existing workflows when integrating LLMs, primarily to ensure appropriate human oversight while maintaining effectiveness and efficiency. Finally, \textbf{R7} contributes to \textit{Technical Robustness and Safety} by promoting knowledge sharing and skill development, which helps practitioners use LLMs more reliably in practice.

The mapping shown in Figure~\ref{fig:fig-5} makes clear that the recommendations align with only \textit{three} of the seven Trustworthy AI principles. The remaining four principles, \textit{Privacy and Data Governance}, \textit{Transparency}, \textit{Diversity, Non-discrimination and Fairness}, and \textit{Environmental and Societal Well-being} are not addressed by the recommendations derived from the case studies reported in Section~\ref{sec:recs}.

The post hoc mapping shows that practices and recommendations based mainly on practical and technical experience do not provide broad regulatory coverage. It suggests that practitioners may focus on visible issues such as control, oversight, and technical reliability, while giving less attention to other Trustworthy AI principles. This points to the need for complementary measures to ensure that the full set of principles is addressed.

\section{Discussion} \label{sec:discussion}

This study provides empirical evidence on the adoption of LLM-based tools to support practitioners in their SE activities within industrial contexts. Our findings show that LLM adoption in SE is inherently multidisciplinary, encompassing aspects of (i) technology adoption, (ii) organisational change management, (iii) human–AI collaboration, and (iv) regulatory compliance. Accordingly, we compare and contrast our findings with those presented in the Related Work (Section~\ref{sec:rw}), relating them to relevant literature across the aforementioned multidisciplinary themes, as presented below.

\subsection{Technology Adoption}

One of the notable findings from both Case 2 and Case 3 is the effect of LLMs on users' experiential learning and their willingness to tackle unfamiliar tasks. This aligns with prior work \citep{Pereira2024,10.1007/978-3-031-42622-3_49}, which reports that LLMs can boost users' learning and confidence when approaching new challenges. Consequently, users are able to engage in a broader range of SE tasks rather than limiting themselves to a few specific use cases.

Findings from Case 2 further suggest that LLM adoption in SE is gradually moving toward a multi-agent approach. LLM-based multi-agent (LMA) systems are found to be better capable of addressing real world challenges that often spread across multiple domains and require expertise from different areas compared to singular LLM-based agents~\cite{10.1145/3712003}, including within SE~\cite{hong2023metagpt}.

Overall, LLM adoption appears to expand practitioners' capabilities, enabling them to tackle a wider variety of tasks while elevating the LLM's role from simply providing suggestions to taking on more agentic responsibilities. This shift also increases the user's accountability and reinforces the need to restructure workflows and maintain robust human oversight.

\subsection{Organisational Change Management}

Some of our findings underscore that successful LLM adoption in SE extends beyond technical integration and requires deliberate change management efforts. The need to restructure workflows (\textbf{R6}) and provide continuous training and knowledge sharing (\textbf{R7}) reflects the principles of organisational change theory, which emphasises readiness, communication, and capability building to sustain transformation~\cite{000243030900014}. The fragmented adoption observed across teams and the call for a more centralised, coordinated strategy (Case 2), along with the confusion caused by the abundance of LLM tools (Case 3) further illustrate the importance of establishing clear vision and leadership. While a grass root level adoption of LLMs is a positive approach that minimises the risk and effect of change resistance, without proper top-down support and leadership strategy, the adoption process will face significant challenges and might not prove to be sustainable. Our findings suggest that LLM adoption should be approached as an organisational change initiative that integrates deliberate strategic planning with the flexibility of an emergent change approach to support long-term success~\cite{10.1002/smj.4250060306}.

\subsection{Human-AI Collaboration}

We found that a majority of participants expressed a strong preference for utilising LLMs as AI assistants rather than mere task automation tools. This preference to perform manual validation of the outputs appears to be tied to concerns about trust and reliability. However, as LLMs become more explainable and their outputs more reliable, it is likely that users' trust in LLMs will increase. This will, in turn, increase their willingness to adopt automation-oriented workflows~\citep{HAQUE2025100204}.

Practitioners also reflected that the helpfulness of the LLM's outputs is prioritised over their accuracy. However, most of the state-of-the-art literature focuses on the quality of the LLM-generated output and benchmark analysis to support their claims~\citep{10449667}. This approach is better suited for evaluating LLM's performance for task automation and aligns with the established theory of conventional human-machine interaction (HMI), which emphasises usability and task performance~\cite{844354}. However, the effectiveness of human-AI collaboration depends not only on system performance but also on the quality of human-AI interaction, trust, and adaptability~\cite{fragiadakis2025evaluatinghumanaicollaborationreview}. These multidimensional aspects are often neglected during evaluations in collaborative workflows, which are a better-suited criteria for task assistance.

Our results are based on practitioners' experiences with LLM-based tools in industrial settings, providing a more relevant context for assessing the feasibility of LLM-based collaborative SE. This reinforces our findings that success from adopting LLM is more closely connected to stakeholder satisfaction with the LLM-assisted workflows than to rigid quantitative benchmarks. 

However, as the degree of LLM automation in SE increases, qualities like accuracy and control becomes increasingly critical, as LLM-based autonomous systems must minimise errors and have fall-back mechanisms in place to maintain user trust and operational reliability~\citep{10.1145/3290605.3300750}. We will need trade-off mechanisms to balance the usefulness and accuracy of LLMs' outputs and the level of control user can maintain over the system.

\subsection{Regulatory Compliance}

Our survey results further reinforce the perceived relevance of our recommendations within various industries that design and develop software products and services. Six out of the seven recommendations received more than 50\% favourable responses (a response of 4 or 5 on the Likert scale) from all survey respondents. \textbf{R4} was the only recommendation that did not.

Recommendation \textbf{R4} is about limiting the application of fine tuned LLMs to use cases for which they were originally intended and designed. This recommendation received only 32\% favourable responses when evaluated for agreement with software practitioners from various industries. However, 69\% of those who did give a favourable response were from the software/IT industry. This indicates that software practitioners from software and IT services industry tend to agree more with \textbf{R4} compared to software practitioners from the other sampled industries.

\textbf{R4} was mapped to the ``technical robustness and safety'' principles from the AI HLEG precisely because the AIA also aims to prevent the misuse of LLMs by disallowing their application beyond their intended scope. However, the lower agreement rate for \textbf{R4} from the sampled practitioners, especially from non-software or IT industries, indicate that either (i) the practitioners are not aware of the impact of the AIA on the adoption of LLMs to assist them in industrial practice, or (ii) they, unwittingly or otherwise, are observed to disregard the regulatory implications of the misuse of LLMs beyond their intended scope. This indicates the need to emphasise the importance and spreading awareness regarding compliance with the AIA and/or any other applicable AI regulations concerning the adoption of LLMs in industrial practice.

An important question raised by our mapping is what it means when empirically grounded, practitioner-driven recommendations align with only a subset of the AI HLEG principles. While one interpretation is that the recommendations simply need to be extended in future work to better cover the remaining principles, another possibility is that this gap could be a reflection of an underlying mismatch between policy-level principles and the realities of day-to-day SE practice. If experienced practitioners do not naturally focus on concerns such as fairness, environmental impact, or transparency when discussing the use of LLMs, in order to be effective, AI HLEG principles must be updated to be sufficiently operationalised, visible, and actionable in real-world development contexts.

\subsection{Novelty Compared to State-of-the-art}

Our findings align with the lessons learnt reported by \citep{Pereira2024}, \citep{10.1007/978-3-031-42622-3_49}, and \citep{WANG2025104113} regarding the necessity of LLM users to possess domain knowledge regarding the task at hand, gaining deeper understanding of LLMs and how they work, limiting the LLM's role to a virtual assistant, and adapting the approach to leveraging LLMs depending on the task.

While prior studies have extensively explored the implications of integrating LLMs into SE contexts, this study introduces several practical, operational-level recommendations that emphasise human oversight, task-specificity, and organisational restructuring. For instance, \citep{Pereira2024}, \citep{10.1007/978-3-031-42622-3_49}, \citep{WANG2025104113} primarily focus on users' learning, collaboration dynamics, and ethical considerations when interacting with LLMs. In contrast, this work provides actionable process-level recommendations for software practitioners seeking to embed LLMs into their workflows in an efficient and responsible manner.

Specifically, the contributions of this study differ in three notable ways: (i) placing emphasis on prioritising usefulness over strict accuracy in LLMs' outputs, (ii) advocating for stakeholder satisfaction as an explicit evaluation criterion for LLM-generated outputs, and (iii) ensuring the selection of appropriate LLM-tools for the task at hand. These perspectives complement existing literature by operationalising higher-level lessons into actionable recommendations for practitioners. Table~\ref{tab:Table-7} summarises how the study's contributions relate to insights from existing works.

\begin{table}[ht]
\footnotesize
\centering
\begin{tabular}{p{5.5cm}p{2.2cm}p{2.1cm}p{2.2cm}}
\toprule
\textbf{Recommendations from this Study} & \textbf{\citep{Pereira2024}} & \textbf{\citep{10.1007/978-3-031-42622-3_49}} & \textbf{\citep{WANG2025104113}} \\
\midrule
Prioritise usefulness over strict accuracy (\textbf{R2}) &  &  & \\ \midrule
Stakeholder satisfaction as evaluation criterion (\textbf{R3}) &  &  & \\ \midrule
Avoid task-misaligned fine-tuned LLMs (\textbf{R4}) &  &  &  \\ \midrule
Restructure business processes to integrate LLMs (\textbf{R6}) &  &  & \checkmark \\ \midrule
Knowledge-sharing activities (seminars, training) (\textbf{R7}) & \checkmark & \checkmark &  \\ \midrule
Use LLMs as assistants with human oversight (\textbf{R1}) & \checkmark & \checkmark & \checkmark \\ \midrule
Implement human oversight and validation mechanisms (\textbf{R5}) & \checkmark & \checkmark & \checkmark \\
\bottomrule
\end{tabular}
\caption{Comparison of this study's contributions with existing related work}
\label{tab:Table-7}
\normalsize
\end{table}

\subsection{Threats to Validity} \label{sec:threats}

We followed Runeson and H\"{o}st's guidelines for qualitative research analysis in software engineering~\citep{runeson2009guidelines} to discuss the potential threats to the validity of this research study.
\newline\newline
\noindent\textbf{Internal Validity}: (1) The selection and grouping of the participants for the ``workshop'' conducted within case 1 is a potential threat as the research team could only advise on that matter. A scrum master from the case company 1 played a crucial role in the design and facilitation of the workshop based on the round-1 interview participants' responses. To mitigate any potential bias arising out of this, the scrum master was informed beforehand of how this would affect the validity of the study and they did not participate in the data collection processes. (2) Despite the promised anonymity, there is a threat of \textit{social desirability bias} being introduced into the responses of the interview participants to provide answers they think align with the organisation's goals rather than their true opinions or experiences. (3) There is a threat of selection bias for the survey participants, especially for the participants recruited via LinkedIn as the fulfilment of the participation criteria is self-reported by the participants and we have no way of verifying that information. (4) Only one author was part of first round of thematic coding performed on the case study data. To mitigate subjectivity bias, we had another author perform coding on a subset of the data, ensuring transparency in the coding process, and incorporating member checking to validate interpretations. However, we did not find any major inconsistencies in the coding and interpretation process between both the researchers and any minor changes or discussions brought up by the second coder was incorporated into the coding process.
\newline\newline
\noindent\textbf{External Validity}: Since the recommendations are synthesised from the findings of the multi-case study, one of the possible threats to the validity of this study is the generalisability of the results. To mitigate this, we conducted a survey with software practitioners from various industries to evaluate the generalisability of the recommendations beyond the context of the case study. (3) Since the data collected from case 1 was based on user experiences of a pilot programme, the short duration of the workshop may have limited participants' ability to fully explore LLM tools. (3) The generalisability of the survey results might not extend to entire population of software practitioners due to potential threat of non-response bias by practitioners who are not proponents of using LLMs.
\newline\newline
\noindent\textbf{Construct Validity}: (1) There is threat of misunderstanding the constructs presented in the interview and online survey by the participants, which might also affect their responses. Since no pilot interviews were conducted within the multi-case study, the understanding of the constructs by the participants from all the cases is not ensured. To mitigate this threat, we employed semi-structured interviews and informed the participants that they were welcome to ask clarification questions during the interview. (2) For the survey, we conducted a pilot survey and follow-up with the respondents of the pilot to ensure the understandability of the constructs within the survey remained consistent across the participants.

\section{Conclusions} \label{sec:conclusion}

In conclusion, our study collected and presented the results of our empirical investigation into the aspects of adopting LLMs for SE activities in industrial settings via a multi-case study. A few of the major findings from the case study focus on aspects like user experiences and perceptions surrounding (i) the preference for LLMs to be used as AI assistants, (ii) the importance of stakeholder satisfaction in the LLM-output evaluation, (iii) scoping the applicability of fine-tuned LLMs, (iv) the effect of LLMs on SE workflows, (v) the necessity and directions for developing human oversight and agency mechanisms, and (vi) necessary skills for practitioners when leveraging LLMs within SE activities. The seven recommendations for the adoption of LLMs within SE activities in practical settings were synthesised based on the findings of our study. To validate the perceived relevance of these seven recommendations, we conducted a survey with software practitioners from various industries. The results of the survey indicate a good level of agreement with most of the presented recommendations from the survey participants. The implications of our post-hoc mapping highlight that assessing AIA compliance only after developing solutions, rather than integrating AIA principles during development, falls short to ensure full compliance. Accordingly, the mapping serves as a contextual``heads-up'' that illustrates where the proposed recommendations already contribute to Trustworthy AI principles, while also indicating where further work is needed for broader compliance.

Building upon the findings of this study, future work should therefore aim to extend the scope of the current recommendations to investigate further the remaining Trustworthy AI principles. In line with the objectives and methodology of this study, such extensions should be empirically grounded and informed by practitioners' perspectives, leading to additional guidance that complements existing recommendations rather than retroactively reshaping them. Since recommendation \textbf{R5}, concerned with the need for human oversight and validation mechanisms, received the highest agreement rate from the survey respondents, researching and developing human oversight mechanisms for employing LLMs responsibly in practical settings is another crucial area of research. Moreover, there is a need for developing further prescriptive artefacts that facilitate the adoption of LLMs into SE activities within industrial settings. As LLMs become more integrated, the evolving role of engineers is an important area of focus. How will their responsibilities shift? What skills and training will be essential? Investigating these questions will be key in guiding the efficient and responsible adoption of LLMs within software organisations. 

\section{Acknowledgements}
This work was supported by the Vinnova project ASPECT [2021-04347] and Wallenberg AI, Autonomous Systems and Software Program (WASP).

\balance
 
\bibliographystyle{elsarticle-harv}
\bibliography{elsarticle-template-harv}

\end{document}